\documentclass[journal]{IEEEtran}

\ifCLASSINFOpdf
   \usepackage[pdftex]{graphicx}
\else
   \usepackage[dvips]{graphicx}
\fi

\usepackage[cmex10]{amsmath}
\usepackage{cite}

\usepackage{url}

\usepackage{setspace}

\makeatletter
\def\url@leostyle{%
  \@ifundefined{selectfont}{\def\UrlFont{\sf}}{\def\UrlFont{\small\ttfamily}}}
\makeatother
\newcommand{\vect}[1]{{\bf #1}}
\newcommand{\mat}[1]{{\bf #1}}

\title{On the Relationship Between Dual Photography \\ and Classical Ghost Imaging }

\author{Pradeep Sen$^\dagger$\thanks{$^\dagger$e-mail: psen@ece.ucsb.edu}\\University of California, Santa Barbara \vspace{-0.18475in}}

\begin{document}

\maketitle

\begin{abstract}

Classical ghost imaging has received considerable attention in recent years because of its remarkable ability to image a scene without direct observation by a light-detecting imaging device.  In this article, we show that this imaging process is actually a realization of a paradigm known as {\em dual photography}, which has been shown to produce full-color dual (ghost) images of 3D objects with complex materials without using a traditional imaging device~\cite{Sen05,Sen09}.  Specifically, we demonstrate mathematically that the cross-correlation based methods used to recover ghost images are equivalent to the light transport measurement process of dual photography. Because of this, we are able to provide a new explanation for ghost imaging using only classical optics by leveraging the principle of reciprocity in classical electromagnetics.  This observation also shows how to leverage previous work on light transport acquisition and dual photography to improve ghost imaging systems in the future.

\end{abstract}



\IEEEpeerreviewmaketitle

\section{Introduction}

\noindent Ghost imaging has been the subject of considerable research within the applied optics community, given its ability to image a scene (usually a 2D partially-transparent slide) without direct observation by a light-sensing imaging device.  Although initially believed to be only a quantum effect~\cite{Pittman95,Abouraddy01}, it was later extended to use pairs of classically correlated photons~\cite{Bennink02,Gatti04a,Gatti04b,Wang04,Cai05,Ferri05}.  However, debate about whether it was intrinsically a quantum effect or classical phenomenon continued (e.g.,~\cite{Erkmen08,Meyers08}), and it was only recently that the optics community accepted that ghost imaging could be performed with fully classical, incoherent illumination~\cite{Shapiro08,Bromberg09}.  

The focus on photon correlation from early ghost imaging work has influenced classical ghost imaging systems even till today, however.  Most systems typically employ a cross-correlation technique between the illumination and the measured value to perform the ghost imaging~\cite{Gatti04a}, although slight variations such as differential imaging~\cite{Ferri10} and normalized ghost imaging~\cite{Sun12} have been proposed.  However, these correlation-based methods all have the fundamental limitation that they are statistical in nature, and so require a large number of measurements ($\sim 10^6$) in order to produce images that approach acceptable quality.  Furthermore, they provide little flexibility in the kind of mathematical tools that can be used to improve their quality or increase their SNR.   To date, there is no published work in the applied optics community that demonstrates full-color ghost images of a 3D scene with complex materials (e.g., scenes with translucent or specular materials, or highly scattering media) with a photographic quality comparable to that of modern digital cameras.

On the other hand, a highly related line of work has been pursued independently by the image-based relighting community in the field of computational imaging.  Specifically, we refer to work on dual photography~\cite{Sen05,Sen09}, which can perform high-quality imaging without using a standard imaging device by modulating the incident illumination to perform imaging.  We note that ghost imaging is effectively equivalent to dual photography, and indeed many of the early results from dual photography resembled the noisy, grayscale images produced by the ghost imaging systems of today. However, dual photography algorithms have improved, and have been shown to produce high-quality, full-color dual (ghost) images of 3D objects with complex materials from the point-of-view of structured light sources, with correct shading effects.

The goal of this paper is to present dual photography within the context of ghost imaging so that the relationship between the two is clear. This will not only allow ghost imaging researchers to benefit from advances in light transport acquisition that have been developed by the image-based relighting community, but it would also allow the two communities to work more closely together in the future.  Although nothing in this paper is actually new (it has all been published before~\cite{Sen05,Sen09}), it accomplishes several important tasks:


\begin{itemize}
\item We demonstrate mathematically the equivalence between the cross-correlation method used in ghost imaging and the light transport measurement of dual photography.

\item We use this equivalence to provide a simple explanation for ghost imaging using the classical electromagnetic principle of reciprocity.  To our knowledge, ghost imaging had not been previously explained in this manner.

\item We show how advances for dual photography and light transport acquisition could be used to improve ghost imaging.  For example, color imaging, efficient acquisition with few light patterns (e.g., less than 1,000), and imaging of real 3D scenes with complex materials have all been demonstrated for dual photography and are therefore possible for ghost imaging.
\end{itemize}

Note that we only focus on ghost imaging using classical, linear optics at macroscopic scales, so incoherent illumination without quantum effects is assumed in our discussion.

\section{Background in Dual Photography}
\label{sec:DualPhotography}

\begin{figure}[t]
\centering
    \includegraphics[width=0.75\linewidth]{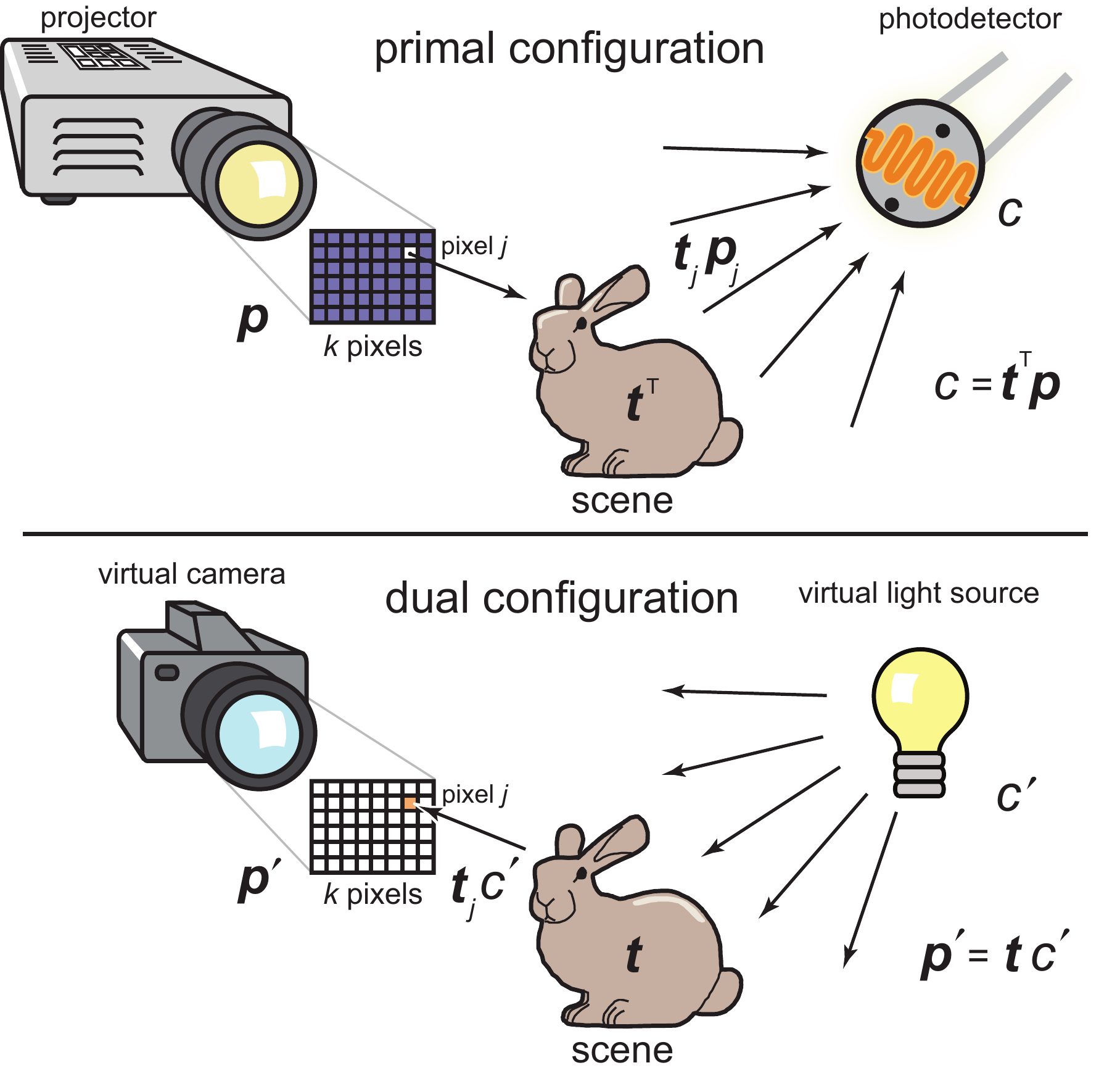}
\vspace{-0.1in}
\caption{\label{fig:PrimalDualConfig}
{\bf Dual photography with single-pixel system.} {\bf (top)} In the primal domain (i.e., the real world) the projector performs the imaging by projecting different illumination patterns.  In this case, for example, a single pixel $j$ is illuminated.  The light arriving at the single-pixel detector $c$ is given by $\vect{t}_j \vect{p}_j$, where $\vect{p}_j$ is the radiant power illuminated by the $j^{\textrm{th}}$ pixel and $\vect{t}_j$ is the transport from this pixel to the detector. {\bf (bottom)} In the dual configuration (done virtually on the computer), the projector is transformed into a camera and the detector into a virtual light source (their respective duals).  If the virtual light source emits radiant power $c'$, then $\vect{t}_j c'$ will be the power arriving through the $j^{\textrm{th}}$ pixel of the virtual camera, accounting for all light paths from the virtual light source to this pixel.  The final dual (ghost) image will be from the point of view of this new camera, while the scene is illuminated by the new light source with correct shading effects.  Images from~\cite{Sen05,Sen05b}.  To see the clear similarity between dual photography and ghost imaging, compare this image with Fig.~S2 in the supplemental materials of~\cite{Sun13}.\vspace{-0.0375in}
}
\end{figure}

\noindent Work on dual photography stemmed from research on image-based relighting~\cite{Debevec00,Masselus03}, where the light transport matrix $\mat{T}$~\cite{Ng03,Sen05} between a set of light sources and detectors was measured for real scenes in order to apply new lighting patterns as a post-process.  We begin by considering the simplest dual-photography system: a single-pixel camera~\cite{Duarte08a} using a projector for structured illumination and a single photodetector to measure outgoing light (Fig.~\ref{fig:PrimalDualConfig}).  This system was first demonstrated by Sen et al.~\cite{Sen05}, although similar systems were later demonstrated by others~\cite{Baheti06,Duarte08b,Sun13}.

Here, the projector's illumination (or speckle field from a spatial light modulator) can be represented by a $k \times 1$ vector $\vect{p}$, where $k$ is the number of independent illumination elements (pixels).  If the single-pixel ``camera'' measures scalar value $c$, the measurement process (assuming classical, linear optics) is $c = \vect{t}^T\vect{p}$, where vector $\vect{t}$ (size $k \times 1$) is the light transport between projector pixels and the detector~\cite{Sen05}.  Specifically, element $\vect{t}_j$ is the fraction of radiant power from projector pixel $j$ measured by the photodetector integrated over all direct/indirect light paths.  The inner-product of $\vect{t}$ and $\vect{p}$ accumulates contributions from all projector pixels and produces measurement $c$.

This framework naturally handles reflection, transparency, and other linear light-transport processes since the only assumption about paths is their linearity.  This is therefore a more general view of the problem than ghost imaging, which has mostly been limited to direct transmission through 2D slides~\cite{Shapiro08,Bromberg09,Katz09,Ferri10,Sun12}.  

Dual photography uses classical electromagnetic reciprocity to virtually exchange the light sources and detectors in a scene~\cite{Sen05}.  In this case, transport $\vect{t}_j$ is reversible: if the detector was a point light source, then $\vect{t}_j$ would also be the fraction of light arriving at the projector through pixel $j$.  In other words, the fraction of the light power arriving at the photodetector from pixel $j$ is also equal to the fraction of the light power arriving at pixel $j$ from a virtual light source positioned at the photodetector.  Note that in this dual domain, the projector is transformed into a camera because projectors angularly emit light while cameras angularly measure light. 

\begin{figure}[t]
\centering
\ifx\ShowTempImages\undefined
    \includegraphics[width=\linewidth]{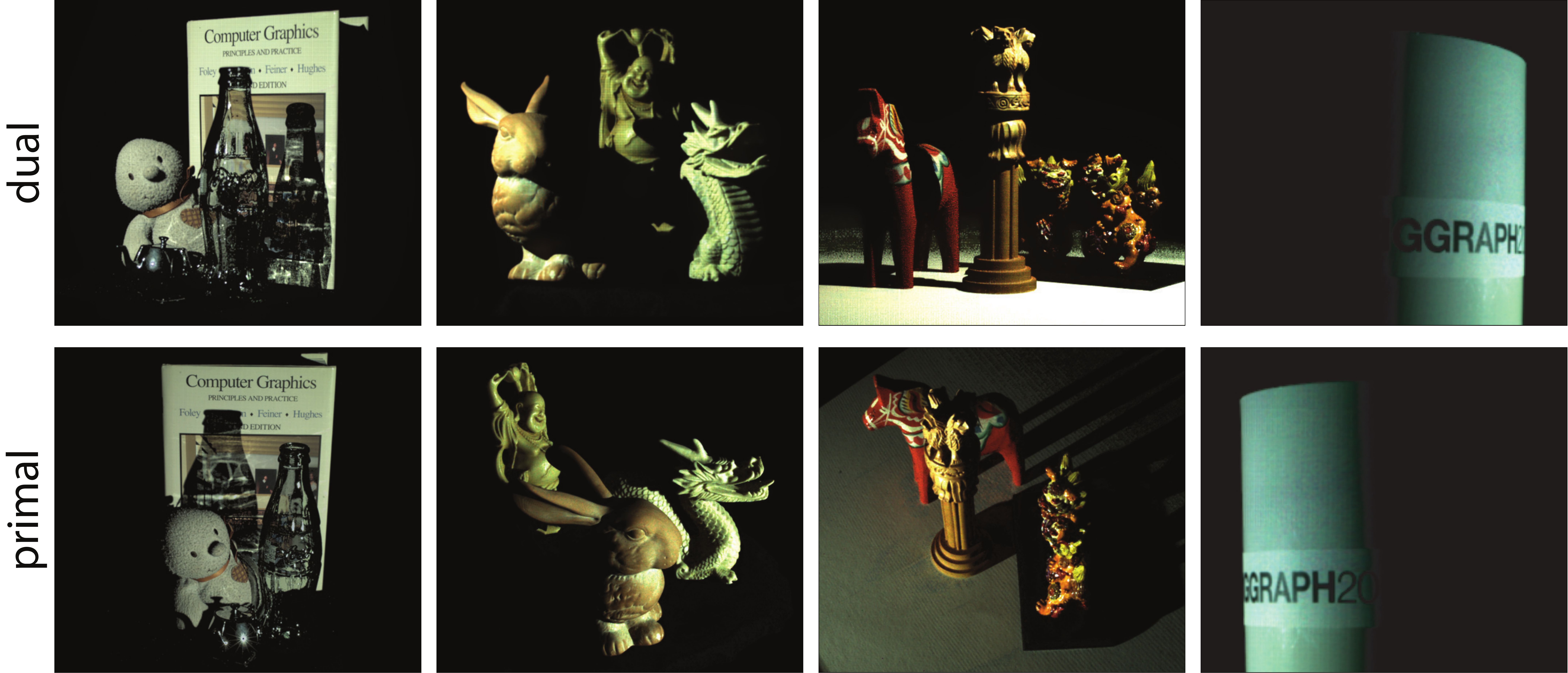}
\else
    \includegraphics[width=\linewidth]{images/DualPhotographsPlaceHolder.eps}
\fi
\vspace{-0.22in}
\begin{small}
  \begin{tabbing}
   ddddddd \=  ddddddddddddd \= ddddddddddddd \= dddddddddddd  \= \kill
   \> {\bf (a)} \> {\bf (b)} \> {\bf (c)} \> {\bf (d)} \\
\end{tabbing}
\end{small}
\vspace{-0.25in}
\caption{\label{fig:SampleDualImages}
{\bf Sample dual (ghost) images, published in 2005~\cite{Sen05}.} The top row shows the dual images reconstructed by dual photography, as seen by the modulated light source which is doing the imaging.  On the bottom row are the primal images, seen from the position of the detector with a fully lit projector.  The first three are natural scenes with 3D objects to demonstrate the capability of dual photography and the last is an experiment to demonstrate that it produces correct shading effects.  {\bf (a)}~Dual photography can handle complex materials, such as translucent materials (the Coke bottle) and specular materials (the metal teapot).  It is interesting that the caustic of the glass bottle in the dual image becomes the actual bottle in the primal and vice-versa.  {\bf (b)}~Scene demonstrating the fairly extreme change of view point that can be done with dual photography.  The objects are seen from the front in the dual image, while the camera sees them from the side.  {\bf (c)}~Dual photography is able to capture details not visible in the primal image.  Note, for example, how the rings in the pillar are clearly visible in the dual image although they cannot be seen in the primal.  {\bf (d)}~Finally, the shading effects are correct based on the new positions of the virtual light sources and imaging devices (see Sec.~\ref{sec:CorrectShading}).  Images reprinted from~\cite{Sen05}.
}
\end{figure}

The $k$-pixel image $\vect{p}'$ that would be ``captured'' by the projector in the dual domain (which we call the dual or ghost image) is given by $\vect{p}' = \vect{t}c'$, where photodetector $c'$ now provides illumination.  Here $\vect{t}$ is the same as before (because of reciprocity), and is proportional to $\vect{p}'$ since $c'$ is a simple scaling factor.  Therefore, the process of dual photography is about measuring the light transport to the photodetector(s) and transposing it to produce an image from the point-of-view of the light source.  Examples of dual (ghost) images are shown in Fig.~\ref{fig:SampleDualImages}.

It is worth noting that the same principle of reciprocity can explain classical ghost imaging as well.  The reason ghost imaging is possible is that there is light transport from the modulated light source doing the ``imaging'' to the simple detector. This transport is subject to the principle of reciprocity and can therefore be reversed, producing a dual (or ghost) image from the point of view of the light source that is ``illuminated'' by the detector.  As we show in the Appendix of this paper, the correlation process at the core of all ghost imaging algorithms is effectively solving for this light transport, which is why it can produce the dual image.

\section{Light Transport Measurement}

\noindent We now examine how the light transport $\vect{t}$ between the light source and the detector is measured, since this is essential for performing dual photography (or ghost imaging).  Conceptually, we apply a set of $n$ light patterns $\vect{c}^T = \vect{t}^T\mat{P}$, where $\vect{c}$ is a vector of $n$ measurements and $\mat{P}$ is a $k \times n$ matrix, and then use our knowledge of $\vect{c}$ and $\mat{P}$ to solve for $\vect{t}$~\cite{Debevec00,Sen05}.

Of course, the incident illumination patterns $\mat{P}$ must be known, which can be easily achieved with digital projectors or spatial light modulators that produce known speckle patterns.  Therefore, imaging systems that use controllable illumination sources (e.g.,~\cite{Sen05,Shapiro08,Sen09,Katz09,Sun13}) require only a single detector.
On the other hand, early classical ghost imaging systems (e.g.~\cite{Bennink02,Gatti04a}) did not know the illumination at $\mat{P}$ {\it a priori} and so they required the explicit measurement of the illumination with a secondary beam which did not interact with the object.   It was the unusual fact that this second beam could be simply correlated with the bucket signal $\vect{c}$ to produce the ghost image that gave ghost imaging a certain air of mystery in the applied optics community.

\begin{figure}[t]
\centering
    \includegraphics[width=0.8\linewidth]{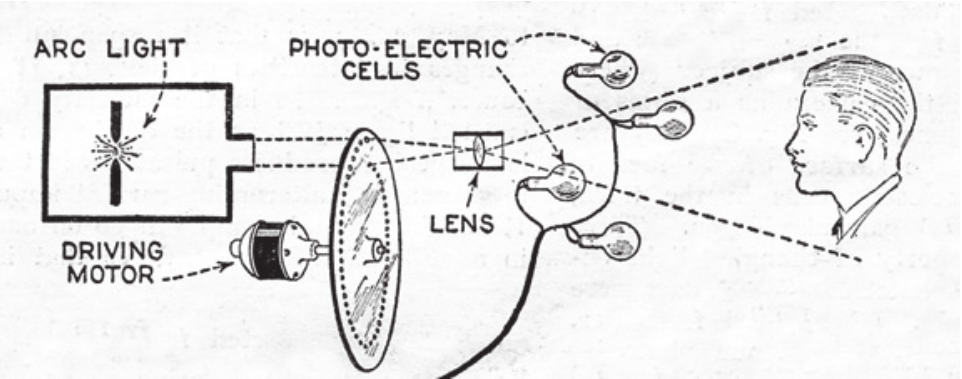}
\vspace{-0.1in}
\caption{\label{fig:FlyingSpotCamera}
{\bf Illustration of flying spot camera, published in 1928~\cite{Baird}.} In this setup, a spinning perforated disk (a Nipkow disk) is illuminated by a strobed arc light to scan the light beam across the scene which is then measured by four photodetectors.  This signal can then be transmitted over the air to a ``television'' which reconstructs the image by modulating a scanning beam based on the measurements.  Note the remarkable similarity between this setup and recent experiments in ghost imaging, such as the experimental setup shown in Fig.~1 of~\cite{Sun13}.  The key technical difference is, of course, that the 1928 version simply scans the illumination beam because random binary patterns were difficult to achieve in a mechanical fashion.  As we show, the approach of \cite{Sun13} which uses binary coded patterns coupled with the correlation-based reconstruction is equivalent to scanning the beam, since they both simply measure the light transport between ``pixels'' in the light source and the photodetectors.  However, the correlation-based approaches typically require more measurements, because the scanning approach only requires as many measurements as there are ``pixels'' in the final image.  The image (courtesy of Wikipedia~\cite{Wiki}) has been edited slightly to fit.
}
\end{figure}

Assuming a properly controllable light source, however, the simplest way to measure $\vect{t}$ is to use an identity matrix for $\mat{P}$, effectively scanning the light beam.  Indeed, scanning beam systems fitted with simple ``backscatter'' detectors produce images from the point-of-view of the scanning source that are ``illuminated'' by the detectors, e.g., Baird's flying spot camera for early television~\cite{Baird} (see Fig.~\ref{fig:FlyingSpotCamera}), or scanning electron microscopes~\cite{Ardenne38}.  This framework shows that these, too, are examples of ghost imaging systems where the ``speckle pattern'' is a known delta function.  When using a digital projector, this effect can be achieved by simply illuminating a single pixel (or a small block of pixels) at a time and scanning over all pixels in the projector to produce an image of the resolution of the projector.  However, these approaches can suffer from low signal-to-noise ratio (SNR)~\cite{Schechner03} and are relatively slow, given that the number of measurements is equal to the effective resolution of the illumination device.

\begin{figure}[t]
\centering
\ifx\ShowTempImages\undefined
    \includegraphics[width=0.24\linewidth]{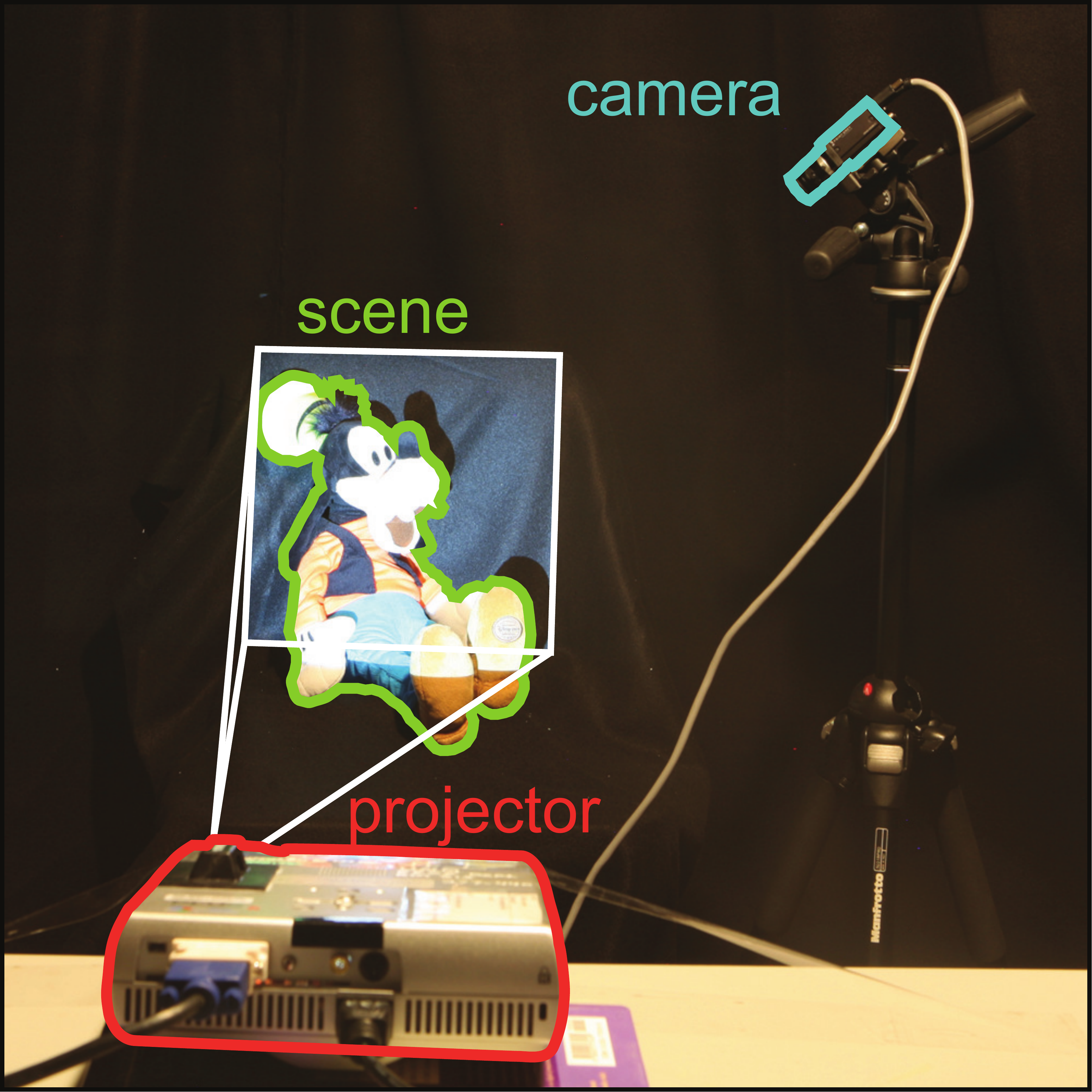}
    \includegraphics[width=0.24\linewidth]{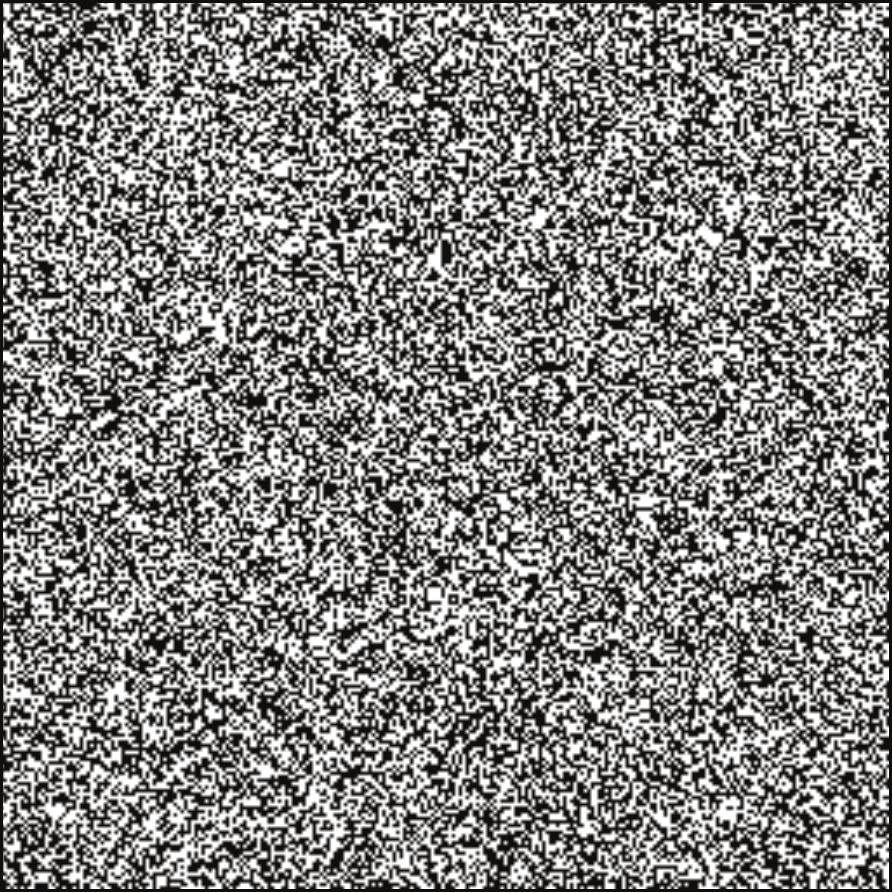}
    \includegraphics[width=0.24\linewidth]{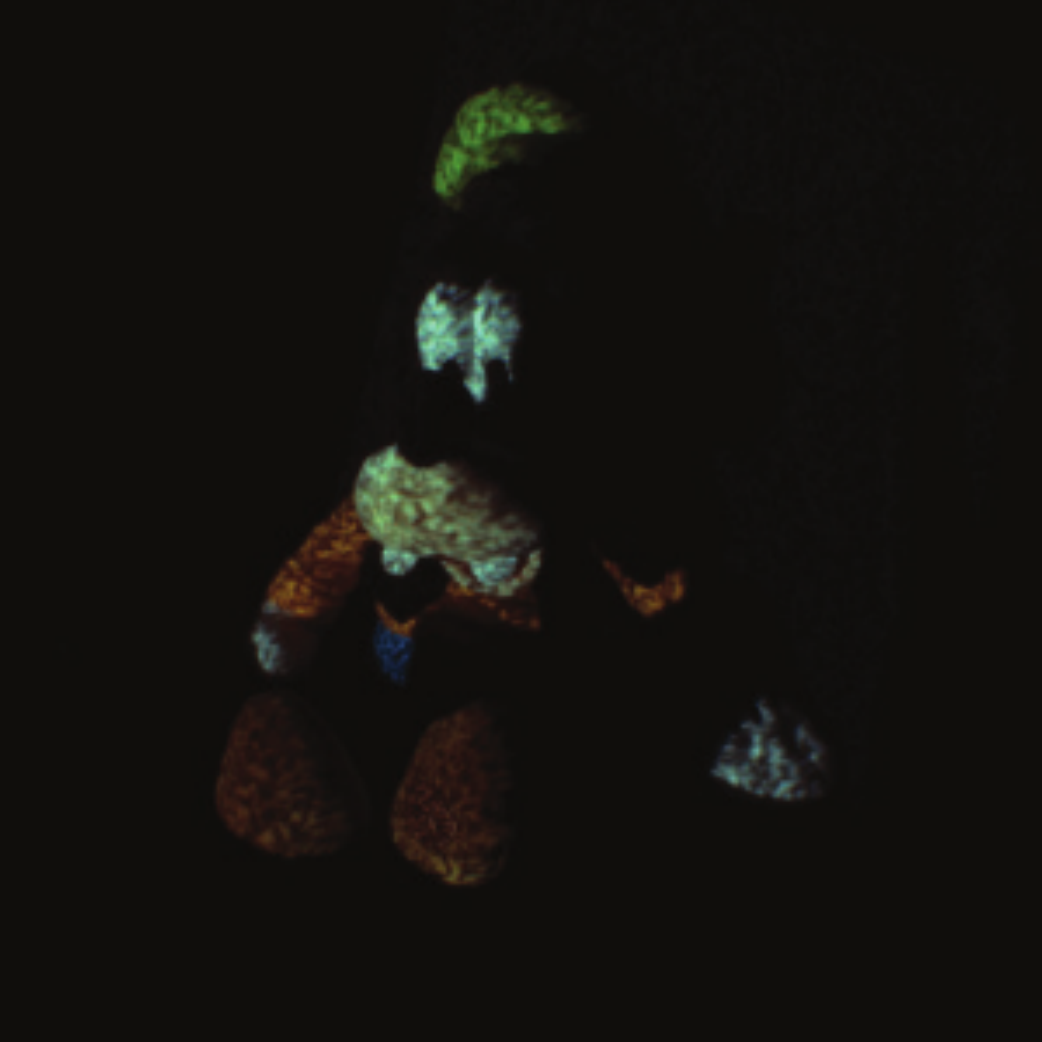}
    \includegraphics[width=0.24\linewidth]{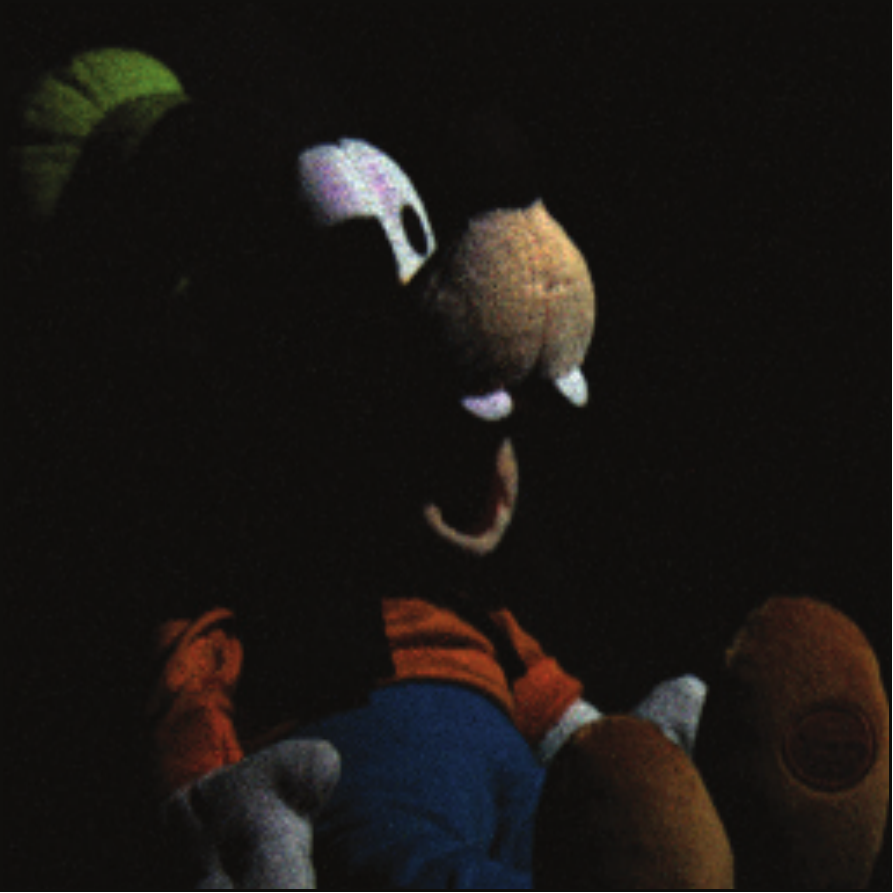}
\else
    \includegraphics[width=0.24\linewidth]{images/OverviewPlaceHolder.eps}
    \includegraphics[width=0.24\linewidth]{images/OverviewPlaceHolder.eps}
    \includegraphics[width=0.24\linewidth]{images/OverviewPlaceHolder.eps}
    \includegraphics[width=0.24\linewidth]{images/OverviewPlaceHolder.eps}
\fi
\vspace{-0.22in}
\begin{small}
  \begin{tabbing}
   dddddd \=  dddddddddddd \= dddddddddddddd \= ddddddddddddd  \= \kill
   \> {\bf (a)} \> {\bf (b)} \> {\bf (c)} \> {\bf (d)} \\
\end{tabbing}
\end{small}
\vspace{-0.3in}
\caption{\label{fig:CompressiveDualPhotographySetup}
{\bf Light transport measurement for compressive dual photography~\cite{Sen09}.}  {\bf (a)} Experimental setup with a digital projector that projects structured illumination patterns onto the scene and a camera (or light detector) to measure the reflected light.  {\bf (b)}~Sample binary light pattern shown by the projector.  In this case the pixels are either black or white with 50\% probability.  {\bf (c)}~Sample image captured by the digital camera above the scene when the pattern in (b) is illuminated. {\bf (d)}~Reconstruction of the dual (ghost) image using compressed sensing~\cite{Candes06,Donoho06}.  The image is of a quality comparable to that of a digital camera with only 1,000 measurements, while the na\"{i}ve scan would have required 65,536 measurements (image resolution: $256 \times 256$).  When compared to recent work in ghost imaging (e.g., Sun et al.~\cite{Sun13}), we see that the acquisition setup is nearly identical.  However, ghost imaging uses correlation-based methods for reconstruction which require many more images ($10^6$) to produce grayscale (not color) results, and these are still visibly noisy and not camera quality. Images reprinted from~\cite{Sen09}.
}
\end{figure}

One can also measure $\vect{t}$ by projecting random patterns and computing the cross-correlation between the detector value and the projected pixels~\cite{Gatti04a}. For example, pixel $j$ in the dual image can be computed as $\vect{p}'_j = \langle (c - \langle c \rangle ) (\vect{p}_j - \langle \vect{p}_j \rangle ) \rangle$ where $\langle \cdot \rangle$ denotes the ensemble average over the $n$ patterns.  As we have discussed, this is the approach typically used in ghost imaging \cite{Ferri10,Sun12,Sun13}. In the Appendix we show that this process effectively measures transport $\vect{t}_j$ since this sets $\vect{p}'_j$ to be equal to $\frac{1}{4}\vect{t}_j$.  Hence, ghost imaging also measures light transport up to a scale, which can be ignored because of factor $c'$ in the dual equation.

This is the reason why the early classical ghost imaging systems (e.g.~\cite{Bennink02,Gatti04a}) could simply correlate the two beams to compute the dual image: one beam was measuring $\vect{p}$ and the other $c$, and their correlation was effectively measuring the light transport between the modulated light source and the detector.  Reciprocity was then implicitly leveraged to produce an image from the point of view of the light that was illuminated by the detector.  The fact that correlated ghost imaging simplifies down to classical light transport measurement also proves it can be a purely classical effect~\cite{Sen05}.

Correlation-based approaches suffer from serious drawbacks, however.  For one, they are statistical methods and so they require a large number of measurements ($\sim \hspace{-0.05in} 10^6$) to produce reasonable results, albeit still visibly noisy.  The observation that classical ghost imaging is equivalent to dual photography enables the use of any approach for measuring light transport (e.g., approaches that are more robust to noise) for improving ghost imaging.  For example, a better approach for measuring the transport is to use an algorithm which adaptively modulates the light patterns in response to measurements to compute $\vect{t}$~\cite{Sen05}.  This is better than cross-correlation in both reconstruction quality and the number of patterns required, although it needs active processing during acquisition.  

An improved method was demonstrated by Sen and Darabi~\cite{Sen09}, which projects simple random binary patterns and uses compressed sensing to reconstruct the light transport (see Fig.~\ref{fig:CompressiveDualPhotographySetup}).  To see how, we rewrite the measurement process by transposing both sides and representing the transport in transform domain $\mat{\Psi}$, yielding $\vect{c} = \mat{P}^T \mat{\Psi} \vect{\hat{t}}$, where $\vect{\hat{t}}$ is the light transport in the transform domain and is assumed to be sparse.  Compressed sensing is then used to solve for the sparsest $\vect{\hat{t}}$ that matches observations $\vect{c}$.  After $\vect{\hat{t}}$ is reconstructed, we can take the inverse transform to get $\vect{t}$, which produces high quality images with a small number of patterns~\cite{Sen09}.

We note that a similar approach using compressed sensing was pursued simultaneously by Peers et al.~\cite{Peers09}, although their system uses a diffuse light source (an LCD display instead of a projector) and hence is only intended to measure light transport and cannot perform imaging.  Hence, although their method is also related to this discussion, it is not directly connected to the application of ghost imaging.  There was also work in the ghost imaging community that attempted to use compressed sensing~\cite{Katz09}, but their approach required much larger percentage of patterns (around 60\%) and was demonstrated using only 2-D scenes of grainy, grayscale slides.  The algorithm of Sen and Darabi~\cite{Sen09}, for comparison, could produce full-color images of photographic quality in some cases with less than 1\% of the total number of measurements.

Finally, there have been other advanced methods for measuring light transport that have been developed in the image-based relighting community which we discuss briefly at the end of the next section.  However, not all of them can be applied to the problem of dual photography (or ghost imaging).

\begin{figure}[t]
\centering
    \includegraphics[width=0.75\linewidth]{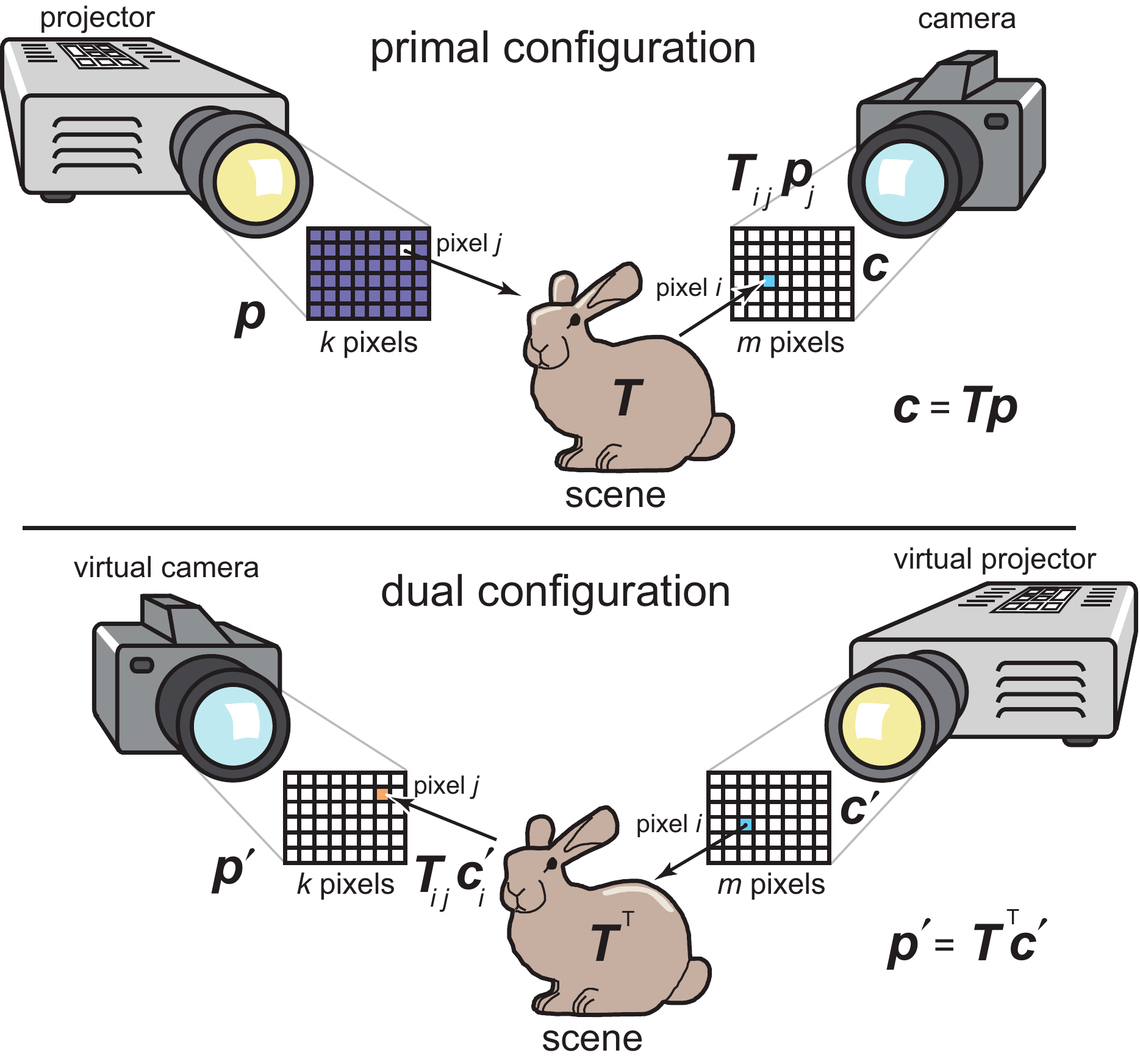}
\vspace{-0.1in}
\caption{\label{fig:PrimalDualConfigWithCamera}
{\bf Dual photography using a camera.} {\bf (top)} If the photodetector in Fig.~\ref{fig:PrimalDualConfig} is replaced by an imaging camera, the system now measures the light transport matrix $\mat{T}$ between the pixels in the projector and the pixels in the camera. By transposing the transport matrix, we get the dual configuration {\bf (bottom)} where the camera has been replaced by a virtual projector.  We can now apply arbitrary ``projector'' patterns at $\vect{c}'$ and get the correct dual image, as shown in Fig.~\ref{fig:RelightingResults}.  Although there is now an imaging device in the setup, this capture process is still equivalent to ghost imaging because the final image is generated from the perspective of the illumination source. Images from~\cite{Sen05}.\vspace{-0.0375in}
}
\end{figure}

\section{Using Cameras for Scene Relighting}

\noindent The dual photography setup described in Sec.~\ref{sec:DualPhotography} can be extended by replacing the photodetector with an imaging camera as shown in Fig.~\ref{fig:PrimalDualConfigWithCamera}.  In this new configuration, the $m$-pixel camera takes measurements while the $k$-pixel projector provides illumination, so the measurement process becomes $\vect{c} = \mat{T}\vect{p}$, where $\vect{c}$ is now an $m \times 1$ vector and matrix $\mat{T}$ (size $m \times k$) represents the light transport between projector and camera pixels~\cite{Sen05}.  Here, element $\mat{T}_{ij}$  represents the fraction of radiant power from projector pixel $j$ measured by pixel $i$ of the camera integrated over all direct/indirect light paths.

In the dual configuration, the projector is virtually replaced by a $k$-pixel camera and the camera by a $m$-pixel projector, with the field of view, focus, and other settings staying exactly the same for each.  The transport in the dual case between the $i^{\textrm{th}}$ pixel of the projector and the $j^{\textrm{th}}$ pixel of the camera is still $\mat{T}_{ij}$ because of reciprocity\footnote{By reciprocity, the light transport between the $j^{\textrm{th}}$ pixel of one device and the $i^{\textrm{th}}$ pixel of another is the same regardless of which way the light is actually flowing.}.  Therefore, the light transport equation from the virtual projector to the virtual camera in the dual domain is given by $\vect{p}' = \mat{T}^T \vect{c}'$.  

\begin{figure}[t]
\ifx\ShowTempImages\undefined
    \includegraphics[width=\linewidth]{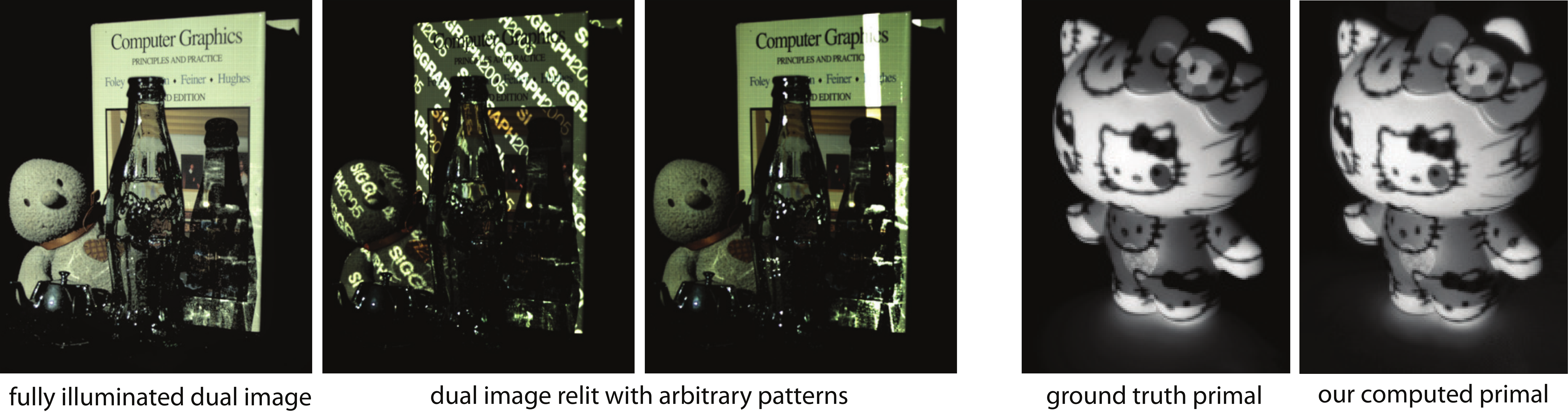}
\else
    \includegraphics[width=\linewidth]{images/RelightingResults_PlaceHolder.eps}
\fi
\vspace{-0.25in}
\caption{\label{fig:RelightingResults}
{\bf Image-based relighting results.} The first set of images shows how the dual image can be virtually relit by applying arbitrary patterns at the virtual projector $\vect{c}'$.  The first of these images is with the virtual projector fully illuminated while the other two project specific patterns.  Note, for example, how the light transport accurately captures the complex caustics through the glass bottle.  The next two images demonstrate the fidelity of the measured light transport matrix using the compressed sensing method of~\cite{Sen09}.  The first is the primal image captured by the camera when the projector is showing a specific pattern (ground truth).  The second is our calculated image by performing the matrix-vector multiply between the light transport matrix and the desired light pattern ($\vect{c} = \mat{T}\vect{p}$).  The two are comparable, indicating that the light transport matrix is accurately representing the flow of light in the scene. Images from~\cite{Sen05} and~\cite{Sen09}.\vspace{-0.0375in}
}
\end{figure}

As with the simple photodetector configuration, this is also an example of ghost imaging since the final image is generated from the point of view of the illumination source without using a standard imaging device at that position.  Like before, the problem reduces to measuring the light transport, in this case matrix $\mat{T}$, which can be done with the methods described in the last section.  Once this light transport matrix has been measured, it can be simply transposed to produce the dual (ghost) images.  Furthermore, we can apply an arbitrary pattern at the virtual projector $\vect{c}'$ to virtually relight the scene as shown in Fig.~\ref{fig:RelightingResults}.  The ability to relight the final image with arbitrary light patterns as a post-process is the reason that the image-based relighting community has focused on measuring the light transport matrix $\mat{T}$, and not the simple light transport vector $\vect{t}$ that ghost imaging systems typically measure.

As mentioned earlier, the relighting community has proposed other advanced methods for measuring light transport more efficiently, such as Symmetric Photography~\cite{Garg06}, the Kernel Nystr\"{o}m method~\cite{Wang09},  optical Krylov subspace methods~\cite{OToole10}, and Primal-Dual coding~\cite{OToole12}.   However, most of these are not applicable to dual photography (and hence ghost imaging) in their current state.  For example, the Kernel Nystr\"{o}m method samples columns and rows of the light transport matrix and reconstructs it assuming that it is low rank in a nonlinear space~\cite{Wang09}.  Measuring columns of the matrix is straightforward (simply illuminate the appropriate pixels in the projector), but sampling the rows is much trickier because it requires a coaxial projector at the camera's position and a coaxial camera at the projector's position to measure the output.  Therefore, these algorithms are only useful for light transport acquisition and not dual (ghost) imaging since they need an imaging camera at the position of the light source, defeating the purpose of ghost imaging.
However, it would be an interesting subject of future work to see if any of these techniques could be modified to work for dual photography and ghost imaging.

\section{Correct Shading in the Dual Image}
\label{sec:CorrectShading}

\noindent Sen et al.\ noted that the shading effects in the dual (ghost) images are correct for the position of the new virtual light sources~\cite{Sen05}, which can be inititally surprising.   Assume, for example, a scene composed entirely of diffuse (Lambertian) surfaces, where the shading is simply proportional to the dot product between the illumination direction and the surface normal.  If the projector is shining orthogonally to a diffuse surface, these regions will be seen as ``bright'' by the detector (photodetector or imaging camera), regardless of its position.
In this case, suppose that the detector is imaging the surface from a grazing angle.  

In the dual domain, the detector would turn into the light source and the projector into the camera.  Since the virtual light is illuminating the diffuse surface at a grazing angle, the surface should be dark.  Is it possible to produce dark values in this region when the detector was measuring bright values before?  The surprising answer is that this is indeed what happens in dual photography (and ghost imaging).

As the proof in Appendix A of~\cite{Sen05} shows, the energy transfer between a pixel of the projector and a pixel of the camera remains the same in either the primal or dual configurations.  However, the shading will change because each camera pixel receives light transport from a different number of projector pixels depending on the configuration.   For example, in the primal domain, a single camera pixel will measure the contributions from many projector pixels because it is viewing the surface at a grazing angle.  So 1 unit of power at each projector pixel will be accumulated over many projector pixels to produce one bright camera pixel.  This will make the Lambertian surfaces with orthogonal light sources appear bright in the primal image.

On the other hand, when the camera is turned into a virtual projector in the dual configuration, each pixel in the virtual projector will map to many dual camera pixels (the same number as before and with the same energy transfer of each).  In this domain, 1 unit of power from a virtual projector pixel will be spread out to many different virtual camera pixels.  Therefore, the results at each pixel of the final image will be much dimmer, as they should be.  Hence the shading will be correct with respect to where the virtual light source is positioned.  This same analysis can be applied to a simple photodetector, which is effectively a 1-pixel camera.  As shown in~\cite{Sen05}, this integration process over different numbers of pixels can be accounted for by adding another cosine term to the light transport equation, which has also used by others, e.g., the Helmholtz Stereopsis work by Zickler et al.~\cite{Zickler02}.  

Experimentally, this has been verified in~\cite{Sen05} using a calibrated experiment, shown here in Fig.~\ref{fig:SampleDualImages}(d).  In this case, the primal projector is almost orthogonal to the first `G' in the ``SIGGRAPH 2005'' logo, so this region appears bright in the primal image.  On the other hand, the `2' is at a grazing angle, so it is dark.  In the dual image, the light source and camera are interchanged, so the `G' is now dark and the `2' is bright.  This shows the shading effects are correct based on the position of the light source.

\section{Comparison to Recent Work in Ghost Imaging}

\noindent Most ghost imaging systems used simple scenes composed of 2D grayscale transparent slides~\cite{Shapiro08,Bromberg09,Katz09,Ferri10,Sun12}, so the relationship with dual photography was difficult to see.  Recently, however, Sun et al.\ made efforts to extend this to imaging 3D diffuse objects~\cite{Sun13}, which made it more clear that this work was a replication of previous work in dual photography.  For example, Sun et al.'s system uses a digital projector to project binary patterns to perform ghost imaging of a diffuse scene with 4 photodetectors.  This setup is similar to previous dual photography work~\cite{Sen05,Sen09}, and the binary patterns and acquisition process are identical to those of compressive dual photography (see Fig.~\ref{fig:CompressiveDualPhotographySetup}).

In their implementation, Sun et al. use a cross-correlation method to reconstruct their dual (ghost) images, which means they need a large number of patterns ($10^6$).  This is inefficient since the resolution of their dual images (which are $158 \times 210$ in Fig.~1 of~\cite{Sun13}) should require at most 33,180 measurements to produce an image~\cite{Sen05}.  With the compressive dual photography method of~\cite{Sen09}, the number of measurements required would be less than 16,000, at least a $60\times$ improvement over the results presented in~\cite{Sun13}.  Furthermore, by following the approaches presented in earlier work~\cite{Sen05,Sen09}, they would be able to produce results in color (either placing color filters in front of the photodetectors or using multicolor illumination patterns) and with less noise.

In the end, Sun et al.'s system produces four dual images from the point-of-view of the projector lit by the different photodetectors, which are then fed into a standard shape-from-shading algorithm to output the final grayscale geometry~\cite{Sun13}.  A drawback of this simple shape-from-shading algorithm is that it requires the surfaces to be entirely diffuse.   For comparison, the image-based rendering community has also used dual photography to reconstruct geometry (e.g.,~\cite{Aliaga09}), but these approaches can reconstruct high-quality, full-color geometry for complex scenes and can handle glossy materials.

\section{Conclusion}

\noindent By showing that ghost imaging reduces to the problem of measuring the light transport between pixels in the projector (or spatial light modulator) and a detector, we make a connection between ghost imaging and previous work in dual photography/light transport acquisition.  This observation will provide researchers in ghost imaging with new tools that could improve their results, such as reducing the number of measurements and producing high-quality, full-color results of scenes with complex materials.
 Although the work to date in dual photography and ghost imaging has been done separately by different research communities, we hope discussions like this will increase their collaboration and lead to improved imaging techniques in the future.

\section*{Acknowledgments}

\noindent We thank our co-authors from the original dual photography papers~\cite{Sen05,Sen09} for valuable discussions.  The compressive dual photography work~\cite{Sen09} was sponsored by NSF CAREER grant \#IIS-0845396.  Images from~\cite{Sen05} \copyright 2005 Association for Computing Machinery, Inc., reprinted by permission.  Images from~\cite{Sen09} reprinted by permission.

\section*{Appendix}

\noindent The following derivation proves that cross-correlation methods used in ghost imaging, which compute final pixels in the form $\vect{p}'_j = \langle (c - \langle c \rangle ) (\vect{p}_j - \langle \vect{p}_j \rangle ) \rangle$ where $\langle \cdot \rangle$ denotes the ensemble average, are effectively measuring light transport:
\vspace{-0.2in}

\begin{align}
	\vect{p}'_j &= \langle (c - \langle c \rangle ) (\vect{p}_j - \langle \vect{p}_j \rangle ) \rangle, \nonumber\\
	 &= \langle c \vect{p}_j - c \langle \vect{p}_j \rangle - \langle c \rangle \vect{p}_j + \langle c \rangle \langle \vect{p}_j \rangle \rangle, \nonumber\\
	 &= \langle c \vect{p}_j \rangle - \langle c \rangle \langle \vect{p}_j \rangle.
	 \label{eq:correlation}
\end{align}
\vspace{-0.2in}

\noindent Assuming random, binary illumination patterns where pixel $\vect{p}_j$ has equal probability being 0 or 1, the expected value $\langle \vect{p}_j \rangle = \frac{1}{2}$.  Given that $c = \vect{t}^T\vect{p}$, we can compute $\langle c \rangle$: 
\vspace{-0.2in}

\begin{align}
	\langle c \rangle = \langle \vect{t}^T\vect{p} \rangle &= \langle \sum_i \vect{t}_i \cdot \vect{p}_i \rangle = \sum_i \vect{t}_i \cdot \langle \vect{p}_i \rangle,  \nonumber\\
	&= \frac{1}{2} \vect{t}^T\vect{1}, \nonumber
\end{align}
\vspace{-0.2in}

\noindent where $\vect{1}$ is a $k \times 1$ vector of all 1's.  A similar calculation can be done for $\langle c \vect{p}_j \rangle$:
\vspace{-0.2in}

\begin{align}
	\langle c \vect{p}_j \rangle &= \langle \sum_i (\vect{t}_i \cdot \vect{p}_i) \vect{p}_j \rangle,\nonumber \\
	&= \langle \sum_{i \ne j} (\vect{t}_i \cdot \vect{p}_i) \vect{p}_j \rangle + \langle \vect{t}_j \cdot \vect{p}_j^2 \rangle,\nonumber\\
	&= \frac{1}{4} \sum_{i \ne j} \vect{t}_i + \frac{1}{2} \vect{t}_j = \frac{1}{4} \vect{t}^T\vect{1} + \frac{1}{4}\vect{t}_j. \nonumber
\end{align}
\vspace{-0.15in}

\noindent Putting this all together into the last line of Eq.~\ref{eq:correlation} gives us:
\vspace{-0.2in}

\begin{align}
	\vect{p}'_j &= \frac{1}{4} \vect{t}^T\vect{1} + \frac{1}{4}\vect{t}_j - \frac{1}{4} \vect{t}^T\vect{1} = \frac{1}{4}\vect{t}_j.
\end{align}
\vspace{-0.2in}

\noindent Hence, performing the correlation method to compute pixel $\vect{p}'_j$ is equivalent to measuring the light transport $\vect{t}_j$ up to a constant $\frac{1}{4}$, which can be ignored because of the $c'$ term in the dual equation.

\setstretch{0.97155}
\bibliographystyle{IEEEtran}
\bibliography{TR_DualPhoto_GhostImaging}
\end{document}